\documentclass[12pt,epsf]{article}
\usepackage{graphicx, amsmath}

\begin{document}

\sloppy
\begin{flushright}{SIT-HEP/TM-36}
\end{flushright}
\vskip 1.5 truecm
\centerline{\large{\bf Cosmological perturbations from}}
\centerline{\large{\bf inhomogeneous preheating and multi-field trapping}}
\vskip .75 truecm
\centerline{\bf Tomohiro Matsuda\footnote{matsuda@sit.ac.jp}}
\vskip .4 truecm
\centerline {\it Laboratory of Physics, Saitama Institute of Technology,}
\centerline {\it Fusaiji, Okabe-machi, Saitama 369-0293, 
Japan}
\vskip 1. truecm
\makeatletter
\@addtoreset{equation}{section}
\def\theequation{\thesection.\arabic{equation}}
\makeatother
\vskip 1. truecm

\begin{abstract}
\hspace*{\parindent}
We consider inhomogeneous preheating in a multi-field trapping model. 
The curvature perturbation is generated by inhomogeneous preheating
 which induces multi-field trapping at the enhanced symmetric point
 (ESP), and results in fluctuation in the number of e-foldings. 
Instead of considering simple reheating after preheating, we consider a
 scenario of shoulder inflation induced by the trapping.  
The fluctuation in the number of e-foldings is generated during this
 weak inflationary period, when the additional light scalar field is
 trapped at the local maximum of its potential. 
The situation may look similar to locked or thermal inflation or even to
 hybrid inflation, but we will show that the present mechanism of
 generating the curvature perturbation is very different from these
 others. 
Unlike the conventional trapped inflationary scenario, we do not make
 the assumption that an ESP appears at some unstable point on the
 inflaton potential. 
This assumption is crucial in the original scenario, but it is not
 important in the multi-field model. 
We also discuss inhomogeneous preheating at late-time oscillation, in
 which the magnitude of the curvature fluctuation can be enhanced to
 accommodate low inflationary scale.  
\end{abstract}

\newpage
\section{Introduction}
\hspace*{\parindent}
The primordial curvature perturbation is supposed to be generated from
the perturbation of some light scalar field, whose fluctuation is
generated during the primordial inflationary expansion.  
In the traditional inflationary scenario the inflaton potential is
supposed to be responsible for both the expansion of the Universe and
the generation of the curvature perturbations. 
Despite the simplicity of the traditional scenario, the scenario in
which the inflaton is responsible both for the inflation and the
curvature perturbation sometimes suffers from serious fine-tunings.  
This problem seems rather evident in low-scale inflationary
models \cite{low_inflation} that will be very important if some
gravitational effect could be observed in LHC.\footnote{In the low-scale
inflationary models there are many problems related to other
cosmological events. Perhaps the most significant condition appears from
baryogenesis \cite{low-baryo, low-AD, low_DG}
 if the scenario of the electro-weak baryogenesis fails.} 
Alternatives to the traditional scenario have been discussed
by many authors\cite{curvaton_1, matsuda_curvaton,
alternate, alt2, Inho_Reh_Dvali, alternate2, SSB-curvaton, 
matsuda_inst1, matsuda_inst2}.
In these alternative scenarios the generation of the curvature
perturbation is mostly due to the late-time conversion of the
isocurvature perturbation that is related to a light scalar field other
than the inflaton. 
The conversion mechanism characterizes the scenario. 
In these alternative scenarios the generation of the seed fluctuations
of the light scalar field occurs during primordial inflation, and thus
the typical length scale of the resultant fluctuation is very large even
if the generation of the curvature perturbation occurs very late.

Among these alternatives an attractive idea of inhomogeneous reheating
was considered by Dvali et al. in Ref.\cite{Inho_Reh_Dvali} and then
extended by many other authors\cite{alternate2}, where
spatial fluctuations in the perturbative decay rate of the inflaton
field to ordinary matter lead to fluctuations in the reheating
temperature. 
Although this idea is very simple and is well motivated by the
moduli-dependent couplings in the string model, we would like to remind
the reader that in actual reheating the so-called preheating would be
efficient before the conventional reheating that will be induced by the
perturbative decay.
Since there are various cosmological scenarios related to the preheating
scenario (only some of which will be discussed in this paper), we think
it is very important to construct a viable scenario for generating the
curvature perturbation that works with preheating.  
This is our motivation to consider the scenario of inhomogeneous
preheating. 
Besides the normal preheating that we will discuss in this paper, another
possibility(instant preheating\cite{instant_original}) has been
considered for brane inflationary model\cite{matsuda_inst1} and 
chaotic inflationary model\cite{SSB-curvaton, matsuda_inst2} 
including MSSM inflation\cite{matsuda_inst2}. 
As we will discuss later in this paper, our present mechanism for
generating the curvature perturbations is quite different from the ones
that have been based on instant preheating, although there is a
similarity in the generation of the preheat field. 
Moreover, we are afraid that the similarity in the names ``inhomogeneous
reheating'' and ``inhomogeneous preheating'' would be a confusing
to the reader even though there is an essential difference 
between the two scenarios.
In the latter scenario (inhomogeneous preheating) the perturbative decay
rate is supposed to be a constant and hence it will play no role in
generating the curvature perturbation, while in the former case the
spatial inhomogeneity in the decay rate plays the essential role.  
In Ref.\cite{matsuda_inst1} we have considered the simplest scenario of
inhomogeneous preheating that induces inhomogeneous expansion during 
trapped inflation, which results in the fluctuation in the number of
e-foldings ($\delta N_e$).
The essence of the inhomogeneous trapping is
very different from the mechanism of inhomogeneous reheating that has
been discussed in Ref.\cite{Inho_Reh_Dvali, alternate2}. 
As we will discuss in Sect.3, inhomogeneous trapping is very natural
in the preheating scenario.
We hope the similarities in these names (``inhomogeneous reheating'' and
``inhomogeneous preheating'') do not create further confusion for our
readers. 

The idea of the present mechanism of inhomogeneous preheating is very
simple. 
We assume that there is a light field($\phi_2$) that is accompanied by
large-scale perturbations during the inflationary stage, supposing that
the fluctuations related to this additional\footnote{This light field is
``additional'' in the sense that it is not responsible for inflation}
 light field can seed the fluctuations
in the number density of the preheat field.
During the preheating stage the kinetic energy of the inflaton($\phi_1$)
is transferred into excitations of the preheat field($\chi$),
which increases the actual number density of the preheat field.
As we will show in Sect.2, the efficiency of the process can be biased
by the expectation value of the secondary field($\phi_2$).
Contrary to the previous analyses related to the trapping scenario, we
do not assume that the ESP appears at some unstable point on the
inflaton potential.
In our present model the ESP is put at the minimum of the inflaton
potential. 
Moreover, instead of considering single-field trapping, we will consider
multi-field trapping in which not only the inflaton but also 
the secondary field is trapped at ESP, where the $\phi_2$-potential
has a local maximum. 
The original assumption that the ESP appears at some unstable point on
the inflaton potential could be very natural in some string models,
while it might be unnatural in other cosmological models. 
Throughout this paper, our argument does not rely on this assumption
although it has played an important role in the original trapping
scenario \cite{beauty_is}.
If the trapping of the field $\phi_2$ induces another inflationary
expansion, although it will be very weak compared with the primordial
inflation, this inflationary stage might look very similar to ``shoulder
inflation'' that appeared in hybrid inflationary model.
The trapping mechanism might also look similar to the thermal trapping
in thermal inflation.  
Moreover, one might think that a similar idea has been discussed in
relation to locked inflation.  
Therefore, we think it is important to show how the present scenario
differs from the others. 
We will shortly review these alternatives and discuss the differences in
an appendix.  
We first describe the basic idea of inhomogeneous preheating and
multi-field trapping in Sec.2 and 3, and then we discuss trapped
inflation and its implication for the curvature perturbation in Sec.4. 
Non-Gaussianity is discussed in Sec.5.

\section{Inhomogeneous preheating}
First we will describe the mechanism of inhomogeneous preheating with
the assumption of multi-field trapping. 
Our discussion is based on estimates of the preheat-field production
given by Kofman et al \cite{Towards_Kof}.
We will consider a model with real scalar fields $\phi_i$ $(i=1,2)$ and
$\chi$, 
\begin{equation}
{\cal L}=\frac{1}{2}\partial_\mu \phi_i \partial^\mu \phi_i + 
\frac{1}{2}\partial_\mu \chi \partial^\mu \chi 
-\frac{g^2}{2}(\phi_1^2 + \phi_2^2)\chi^2
-V_i(\phi_i),
\end{equation}
where a real inflaton $\phi_1$ and an additional light scalar $\phi_2$
interact with a preheat field $\chi$.
We will assume that the inflation is a chaotic type and the inflaton
potential is given by
\begin{equation}
V_1(\phi_1) = \frac{\lambda_1|\phi_1|^{n_1}}{M_1^{n_1-4}},
\end{equation}
and the additional light field has a flat potential
\begin{equation}
V_2(\phi_2) = -\frac{1}{2}m^2 \phi_2^2 +
\frac{\lambda_2|\phi_2|^{n_2}}{M_2^{n_2-4}},
\end{equation}
which has a local maximum at the origin $\phi_2=0$.
We will consider the case where the inflaton $\phi_1$ approaches the
origin and generates the preheat field $\chi$ through preheating.
The preheat field might instantly decay into light
fermions by the interaction $\sim g_\chi\chi \bar{\psi}\psi$
if the coupling $g_\chi$ were strong, but here
we assume that $g_\chi$ is small enough to prevent the instant
reheating or is even absent in the model. 
The mass of the preheat field $\chi$ depends on both $\phi_1$ and
$\phi_2$ and is given by
\begin{equation}
m_\chi(\phi_1,\phi_2)= g \sqrt{\phi_1^2+\phi_2^2}.
\end{equation}
Thus, immediately after the end of chaotic inflation, when $\phi_1\sim
M_p$, the mass of the preheat field $m_\chi$ is very large 
provided that $g$ is not extremely small.
In this paper we will assume $g\sim 1$. 
Here we assume that the mass of the light scalar $\phi_2$ is not so 
heavy that it cannot start oscillation immediately after inflation.
The adiabatic condition is violated when 
$|\dot{m}_\chi|/m_\chi^2 \sim |\dot{\phi}_1|/g (\phi_1^2+\phi_2^2) >1$,
where particle production occurs.
Thus the nonadiabatic condition is given by
\begin{equation}
|\phi_1| < \phi_1^*\equiv \sqrt{\frac{v}{g}},
\end{equation}
where $v$ denotes the absolute value of the inflaton velocity
 near the origin.\footnote{As we will discuss later in this section,
 $\phi_2$ is assumed to be small so that it does not suppress the
 efficient production of the preheat field.}
Obviously the nonadiabatic region is very narrow compared with the initial
amplitude of the inflaton $\bar{\phi}_1\sim M_p$.
As a result, we assume that the efficient particle production
occurs almost instantaneously within the time interval 
$\Delta t_* \sim \phi_1^*/v\sim (vg)^{-1/2}$.
Since this time interval is much smaller than the age of the Universe,
 one may neglect the expansion of the Universe during the particle
 production. 
Integrating over the momenta of the preheat field, the number density of
the preheat field that is produced at the first impact is
obtained \cite{Towards_Kof}, 
\begin{equation}
\label{preh_n}
n_\chi = \frac{(gv)^{3/2}}{(2\pi)^3}\exp\left(-\frac{\pi m_\chi^2}{gv}
\right).
\end{equation}
In the multi-field trapping, the initial value of the light field
$\phi_2$ gives the effective mass to the preheat field at the first
scattering.
In this case the value of the effective mass is supposed to be   
small so that it does not suppress the efficient preheating.
The required condition is
\begin{equation}
\label{pre-cond}
\frac{\pi m_\chi^2 }{gv}\simeq \frac{\pi g \phi_2^2}{v} <1.
\end{equation}
The successive scattering is discussed in Ref.\cite{Towards_Kof}
which leads to the equations describing the
occupation numbers of the preheat field with momentum $k$ that is
produced when the inflaton field passes near the ESP for $j+1$ times
as
\begin{equation}
n_k^{j+1} =  b_k^j n_k^j
\end{equation}
where 
\begin{equation}
b_k^j =1+2e^{-\pi \mu^2}-2\sin \theta^j e^{-\pi \mu^2/2}\sqrt{
1+e^{-\pi \mu^2}}.
\end{equation}
Here $\mu^2=(k^2+m_\chi^2)/gv$, and $\theta^j$ is a relative phase that
changes almost randomly if the time interval during each scattering is
not very much shorter than the typical time scale for the change of the
system parameters.

In the above calculation of the preheating, the primordial
fluctuations related to
the light field $\phi_2$ appears in the 
fluctuation of the effective mass $m_\chi$.\footnote{In Ref.\cite{alt2},
Vernizzi has discussed another scenario for generating cosmological
perturbations with mass variations.}
Then the fluctuation that is produced at the first scattering is
\begin{equation}
\label{fluc_pre}
\frac{\delta n_\chi}{n_\chi}\simeq\frac{2\pi g \phi_2 \delta \phi_2}{v},
\end{equation}
and the fluctuations that are generated at successive scatterings are
\begin{equation}
\frac{\delta b^j}{b^j}\sim \frac{2\pi g \phi_2\delta \phi_2}{v}.
\end{equation}
As we will discuss in the next section, soon after the first scattering
$\phi_2$ is trapped at the ESP and starts to
oscillate with much smaller amplitude than its initial value.
The value of the velocity $v$ decreases with time because of the
damping effect, but its rate of decrease cannot overcome the sudden
decrease of $\phi_2$.
In this sense the fluctuations generated from $\delta b^j$
 cannot overcome the one generated at the first scattering.
As a result, the fluctuation generated at scattering is largest
at the first scattering. 
Hence, although the scatterings occur many times during preheating,
it would be fair to assume that the result is always given by
Eq.(\ref{fluc_pre}).

\section{Multi-field trapping}
\begin{figure}[ht]
 \begin{center}
\begin{picture}(410,310)(0,0)
\resizebox{14cm}{!}{\includegraphics{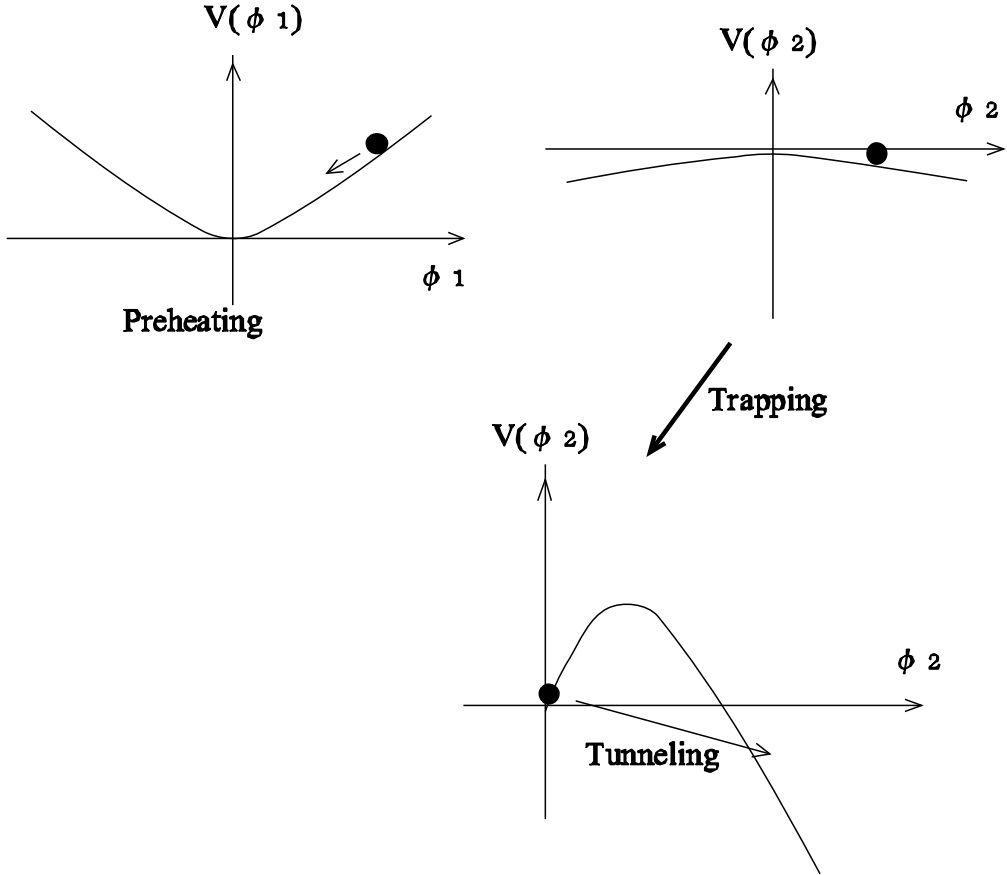}} 
\end{picture}
\caption{The potentials for $\phi_1$ and $\phi_2$ are shown in the first
  and the second picture. Preheating occurs due to the
  $\phi_1$-oscillation, while the trapping occurs for both fields.
  Since the field $\phi_2$ gives the mass for the preheat field $\chi$ 
  at the minimum of the $\phi_1$-potential, the fluctuation 
  $\delta \phi_2$ will induce inhomogeneous 
  preheating. 
  The $\phi_2$-potential during trapped inflation is shown in
  the third picture. 
  Since the potential barrier $\Delta V$ decreases as $\Delta V \propto n_\chi^2$, trapped inflaton ends by the $\phi_2$-tunneling.}
\label{figure1}
 \end{center}
\end{figure}

In Fig.\ref{figure1} we show the potentials for $\phi_1$ and $\phi_2$.
$\phi_1$-trapping and the tunneling is shown in the third picture.

The idea of the trapping mechanism at the ESP has been discussed by Kofman
et al. in Ref.\cite{beauty_is}.
As we have seen above, the motion of the inflaton $\phi_1$ near the ESP 
induces preheating.
During preheating some of the kinetic energy of the inflaton will be
transferred into excitations of the preheat field $\chi$, which finally
leads to the number density of the preheat field $n_\chi \sim
v^{3/2}g^{-1/2}$ \cite{beauty_is}.  
As the inflaton passes away from the ESP the mass of the preheat
field increases, which leads to another channel for the energy
transfer from the inflaton to the preheat field.
In this case the energy density of the preheat field is given by
\begin{equation}
\rho_\chi \simeq g |\phi_1| n_\chi,
\end{equation}
which grows as the inflaton goes away from the ESP.
At this point the backreaction of the preheat field adds an effective
confining potential to the inflaton $\phi_1$, and then the amplitude of
the $\phi_1$-oscillation becomes much smaller than the original value.
This is the basic idea of the trapping after preheating. 

After the first scattering the light field $\phi_2$ starts to feel the
strong confining potential $\sim g\phi_2 n_\chi$, and then starts to
oscillate around the ESP with a very small amplitude.
Therefore, although the roles played by the two fields ($\phi_1$ and
$\phi_2$) in the
inhomogeneous preheating scenario are completely different, they are
both trapped just after the first scattering. 

This is the idea of the multi-field trapping.
We will use this idea to describe the late-time generation of
 the curvature perturbation.

\section{$\delta N_e$ from trapped inflation after inhomogeneous preheating.} 
We discussed above how inhomogeneous preheating and multi-field
trapping occur at the ESP.
Perhaps the simplest scenario for generating the curvature perturbation
with the inhomogeneous preheating would be the instant preheating
scenario, in which the preheat field is supposed to decay with a
homogeneous mass and a homogeneous decay rate when it (nearly) dominates
the energy density of the Universe\cite{instant_original}. 
In this case the fluctuation $\delta n_\chi$ will be 
converted into the fluctuation of the reheating
temperature\cite{matsuda_inst1, matsuda_inst2}. 
Instead of considering the instant preheating scenario, here
we will consider a
scenario of trapped inflation that leads to the late-time generation of
$\delta N_e$, the fluctuation of the number of e-foldings\footnote{See
also Ref.\cite{alternate}}.
We think it will be straightforward to see that the typical length 
scale of the spatial fluctuation that is related to $\delta N_e$ is not 
determined by the expansion during the trapped inflation, but is 
determined by the length scale of the primordial $\phi_2$-fluctuation
that has been generated during the primordial inflation.
We show a schematic picture of the $\delta N_e$ generation in Fig.2.

\begin{figure}[ht]
 \begin{center}
\begin{picture}(410,210)(0,0)
\resizebox{14cm}{!}{\includegraphics{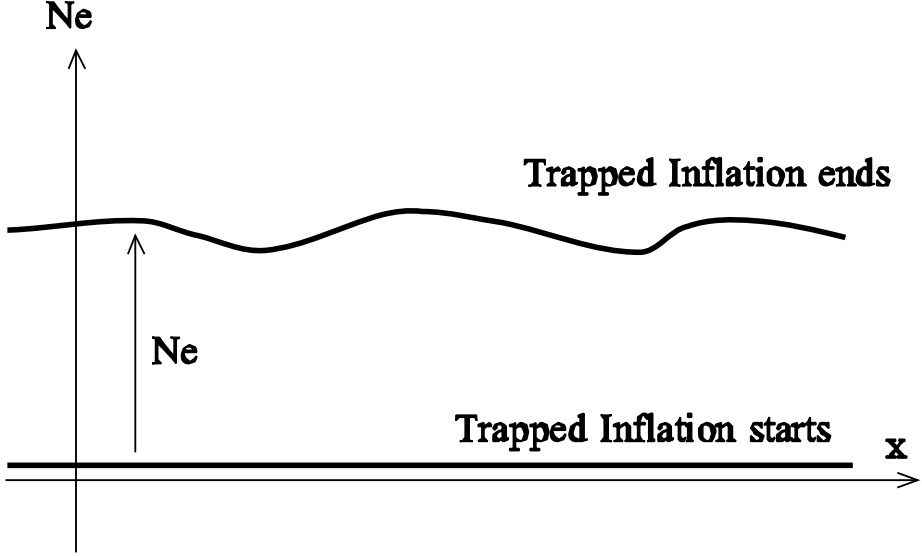}} 
\end{picture}
\caption{The start-line of the trapped inflation is independent of
  the fluctuation $\delta n_\chi$ and is given by the flat
  surface (the straight line at $N_e=0$). On the other hand, the
  end-line is determined by the number density of the preheat field
  $n_\chi$, which has the fluctuation $\delta n_\chi$ induced by the
  primordial fluctuation $\delta \phi_2$. Note that $\delta \phi_2$ has
  left the horizon  during the primordial inflation.}
\label{figure2}
 \end{center}
\end{figure}

Now, let us consider the case where the vacuum energy around the ESP
(i.e., 
the potential at the local maximum of $\phi_2$) dominates the energy
density of the Universe during some period of time after preheating.
This is a multi-field version of the original idea of trapped inflation
that has been discussed in Ref.\cite{beauty_is} and \cite{matsuda_inst1}.
During trapped inflation the effective potential for the light scalar
$\phi_2$ is given by 
\begin{equation}
V_2^{eff}(\phi_2) = V_0 -\frac{1}{2}m^2 \phi_2^2 +
\frac{\lambda_2|\phi_2|^{n_2}}{M_2^{n_2-4}} + g n_\chi |\phi_2|,
\end{equation}
where the constant potential $V_0$ must be tuned so that the value of
the present cosmological constant does not exceed the observational
bound.\footnote{In order to avoid the serious domain-wall problem, we
need to include a $Z_2$-breaking parameter. 
The cosmological problem of 
the domain walls in supersymmetric model can be solved very
naturally \cite{matsuda_wall} if the symmetry of the domain wall is
related to the R-symmetry.} 
Here the mass $M_2$ is supposed to be much larger than $m$.
Looking at the effective potential near the origin, the effective
potential for $\phi_2>0$ is written as
\begin{equation}
V_2^{eff}(\phi_2) \simeq V_0 -\frac{1}{2}m^2 \left(\phi_2 - \frac{g n_\chi}
{m^2}\right)^2 +\frac{g^2 n_\chi^2}{2m^2}.
\end{equation}
Note that both $\phi_1$ and $\phi_2$ are strongly trapped at the origin
just after the preheating because of the large number density of the
preheat field ($n_\chi \sim v^{3/2}g^{-1/2}$).
Then the trapped inflation will be terminated when $n_\chi$ decreases
with time and finally the tunneling from $\phi_2=0$ to $\phi_2 > 2\Delta
\phi_2 \equiv 2 g n_\chi/m^2$ occurs. 
The rough estimate of the tunneling rate \cite{tunnel_dun} suggests
that the tunneling occurs when
\begin{equation}
\label{trap-end}
B \sim \frac{(\Delta \phi_2)^4}{\Delta V} \sim 1,
\end{equation}
where $\Delta V$ is the height of the potential barrier,
\begin{equation}
\Delta V \simeq \frac{g^2 n_\chi^2}{2m^2}.
\end{equation}
Therefore, the trapped inflation will be terminated when $n_\chi$ is
diluted down to $n_\chi < m^3/g$, where the tunneling occurs.
Then the number of e-foldings elapsed during the trapped inflation is 
given by
\begin{equation}
N_e \sim \frac{1}{3}\ln\left(\frac{n_\chi(t_i)}{n_\chi(t_e)}\right).
\end{equation}
In the present model, $n_\chi(t_{i})$ at the beginning of trapped
inflation is supposed to be fluctuated according to Eq.(\ref{fluc_pre}),
while $n_\chi(t_e)$  
at the end of inflation is determined by Eq.(\ref{trap-end}).
Assuming that the primordial curvature perturbation is negligible, 
and also that the fluctuation related to the number of the preheat field
does not induce curvature perturbation before the trapping;
one can see that the curvature perturbation 
$\zeta =\delta N_e \sim \frac{1}{3}\frac{\delta n_\chi(t_i)}
{n_\chi(t_i)}$ is generated by trapped inflation.
Since the fluctuation $\delta N_e$ is generated 
due to the primordial fluctuation related to the light field 
$\phi_2$,\footnote{Note that this fluctuation is not generated during
trapped inflation but has been generated during the primordial
inflation} the typical length scale of
 $\delta N_e$ is very large
 compared with the expansion of the Universe due to trapped inflation.
There might be another possibility that the process of the inhomogeneous
preheating itself could become the origin of the curvature perturbation.
It is very hard to completely exclude this possibility without using
numerical calculations; however there is a reason that we can believe
the generation does not occur during preheating.
The reason is that during the oscillatory regime the energy
density of each field will evolve as $\sim a^{-4}$ \cite{dimo_6223}
and hence the total energy density will evolve as $a^{-4}$ irrespective
of the isocurvature fluctuation related to $\delta
n_\chi$.\footnote{Note that $m_\chi \propto a^{-1}$ and $n_\chi \propto
a^{-3}$ during this regime\cite{dimo_6223}. }
Therefore, the energy density at the beginning of trapped inflation is
determined by $V_0$, while the end-point of trapped inflation is
determined by $n_\chi(t_e)\simeq m^3/g$.
Since the isocurvature perturbation related to
$\delta n_\chi$ exists at the beginning of trapped inflation,
 the total $N_e$ elapsed during trapped 
inflation is fluctuated as we have described above.
Finally the condition for the spectrum of the perturbation is given by
\begin{equation}
{\cal P}^{1/2}_\zeta \sim \frac{ g\phi_2 H_I}{v}\sim 
\frac{g\phi_2}{M_p} \simeq 10^{-5}
\end{equation}
for $v\simeq H_I M_p$, 
which determines the initial condition for the light field $\phi_2$
at the beginning of the preheating.
Since $\phi_2$ is bounded from above by the condition
(\ref{pre-cond}), the condition for $H_I$ is given by
\begin{equation}
\sqrt{\frac{g H_I}{M_p}}>10^{-5}.
\end{equation}
It is possible to consider a hybrid-type inflationary model 
in which $V_0$ dominates the energy density during the primordial
inflation and hence the velocity of the inflation near the ESP will be
much smaller than the above value.
Introducing a small parameter $\epsilon$ that is defined as
$v = \epsilon H_I M_p$, we obtain for the hybrid-type inflationary
model 
\begin{equation}
{\cal P}^{1/2}_\zeta \sim \frac{ g\phi_2 H_I}{v}\sim 
\frac{g\phi_2}{\epsilon M_p} \simeq 10^{-5},
\end{equation}
which leads to the condition
\begin{equation}
\label{preeq}
\sqrt{\frac{g H_I}{\epsilon  M_p}}>10^{-5}.
\end{equation}
Besides the above lower bounds for $H_I$,
in both cases there is an upper bound for $H_I$ depending on the
potential for the primordial inflation, since our basic 
requirement is that the the primordial curvature perturbation that could
be generated by the traditional inflationary mechanism does not dominate
the cosmological perturbation.
For example, in chaotic inflationary model with quadratic potential, 
we obtain the bound
\begin{equation}
\label{prev-H}
10^{13} GeV > H_I > \frac{\epsilon}{g}10^{8}GeV,
\end{equation}
which suggests that small coupling constant $g< \epsilon 10^{-5}$ is not
acceptable in our model.

\section{Inhomogeneous preheating and late-time oscillation}
Inhomogeneous preheating may occur whenever oscillation starts during
the evolution of the Universe.
Here we will consider the case where $\phi_1$ is a light field that
starts to oscillate late after the inflaton oscillation but before the
$\phi_2$-oscillation, taking that 
the $\phi_1$-velocity at the first scattering at the ESP is given by
$v\simeq H(t_{osc})\phi_1(t_{osc})$,
where $t_{osc}$ is the time when $\phi_1$ starts to oscillate.
Then what happens after the first scattering is almost the same as in the
previous scenario provided that the other light field $\phi_2$ satisfies
the required condition $g \phi_2^2/v <1$.
The trapped inflation starts when $V_0$ starts to dominate the energy
density of the Universe, and ends when $n_\chi$ decreases
to $n_\chi \simeq m^3/g$.
Here the form of the $\phi_2$-potential is supposed to be the same as
the previous one. 

The number density of the preheat field generated by the preheating is
$n_\chi \sim v^{3/2} g^{-1/2}$ \cite{beauty_is}.
If the energy density of the Universe evolves as
$\propto a^{-4}$ before trapped inflation,
$n_\chi$ at the beginning of the trapped inflation is
\begin{equation}
n_\chi \sim \frac{v^{3/2}}{g^{1/2}} \times 
\left(\frac{V_0}{H^2(t_{osc})M_p^2}\right)^{3/4}.
\end{equation}
Demanding that the number density of the preheat field at the beginning
of trapped inflation is larger than $n_\chi\simeq
m^3/g$, we find the condition 
\begin{equation}
\phi_1 > \frac{m^2 M_p}{g^{1/3}V_0^{1/2}}.
\end{equation}
The condition for the spectrum of the perturbation is given by
\begin{equation}
{\cal P}^{1/2}_\zeta \sim \frac{ g\phi_2 H_I}{v}
\simeq \frac{ g\phi_2 H_I}{H_{osc}\phi_1} \simeq 10^{-5},
\end{equation}
which is a much looser condition than the one obtained in the previous
scenario 
because of the possible hierarchy $H_I/H_{osc}\gg 1$ and 
$\phi_1/M_p \ll 1$.
The situation looks very similar to the curvaton paradigm in which
the additional phase transition accommodates low inflationary
scale\cite{matsuda_curvaton}. 
Since the dimensionless parameter $\epsilon$ is now given by
\begin{equation}
\epsilon = H_{osc}\phi_1/H_I M_p,
\end{equation}
the condition (\ref{preeq}) obtained in the previous section leads to 
\begin{equation}
H_I > 10^{-5} \sqrt{\frac{H_{osc}\phi_1}{g}}.
\end{equation}

\section{Non-Gaussianity}
In the above analysis we neglected higher terms that are
proportional to $(\delta \phi_2)^n$ ($n\ge 2$).
This approximation is not appropriate if the impact parameter is small
while the fluctuation of the impact parameter is large.  
Therefore, we will examine the non-Gaussianity condition for more
details and see how we can estimate the non-Gaussianity parameter
$f_{NL}$. 
The value of the non-Gaussianity parameter $f_{NL}$ is determined by the
non-Gaussian contribution to the Bardeen potential 
\begin{equation}
\Phi = \Phi_G + f_{NL}\Phi_G^2,
\end{equation}
where $\Phi_G$ denotes the Gaussian part.
The relation between $\Phi$ and $\zeta$ is given by
\begin{equation}
 \Phi = -\frac{3}{5}\zeta =-\frac{1}{5}\frac{\delta n_\chi}{n_{\chi}}.
\end{equation}
The expression for $\delta n_\chi / n_\chi$, which is approximately
given by Eq.(\ref{fluc_pre}) contains higher terms that are
proportional to $(\delta\phi_2)^2$.
The expression for $\delta n_\chi / n_\chi$ up to the second order is
given by
\begin{equation}
\label{2ndorder}
\frac{\delta n_\chi}{n_\chi} \simeq
-\frac{2\pi g \phi_2 \delta \phi_2}{v} 
-\frac{1}{2}\left(\frac{2\pi g}{v}
-\frac{4\pi^2 g^2 \phi_2^2}{v^2}
\right)(\delta \phi_2)^2.
\end{equation}
Therefore, the non-Gaussianity parameter is
\begin{equation}
-\frac{3}{5}f_{NL} \simeq
\frac{3v}{4\pi g \phi_2^2} 
-\frac{3}{2},
\end{equation}
where the factor of $3/2>1$ appears.
Therefore, unlike the traditional inflationary scenario 
the non-Gaussianity parameter $|f_{NL}|$ in our present model is
always\footnote{Here we do not consider fine-tunings that may result in
the cancellation between the two terms.} larger than unity.

\section{Conclusions and Discussions}
\hspace*{\parindent}
\label{sec:conclusion}
In this paper we described an alternative to the traditional
inflationary scenario using the ideas of 
(1) multi-field trapping, (2) inhomogeneous preheating and (3) trapped
inflation. 
As we have discussed in Ref.\cite{matsuda_inst1}, it is
straightforward to apply this mechanism to the brane 
inflationary model. 
The curvature perturbation is generated not by the
primordial inflation but by the trapped inflation,
while the isocurvature
fluctuation that seeds the generation of the curvature perturbation at a
later stage is generated during primordial inflation. 
In the present
case the non-Gaussianity parameter $f_{NL}$  is always larger than
unity. 
Such a large non-Gaussianity is a distinguishable feature of the model.

\section{Acknowledgment}
We wish to thank K.Shima for encouragement, and our colleagues at
Tokyo University for their kind hospitality.
\appendix
\section{Other models}
\begin{itemize}
\item Locked Inflation\\
Note that if there were direct coupling 
$\sim g^2\phi_1^2 \phi_2^2$, the model turns into a conventional hybrid
      scenario.
Then the light scalar $\phi_2$ will be trapped at the origin during
      inflation and also during the inflaton oscillation. 
This is a scenario called ``locked inflation'' that occurs after hybrid
      inflation, and would be very important if one is considering a
      hybrid 
      inflationary model. 
However, in this scenario one cannot generate the curvature perturbation
from the fluctuation related to the light scalar $\phi_2$, simply because
the light field grows to a huge mass $\sim g \phi_1$ during inflation.

\item Thermal inflation\\
We will also comment on the thermal inflationary model, which has a
similar flat potential and is also based on the similar idea that a light
      scalar is trapped at the origin to trigger weak inflation.
Obviously the crucial difference between the two scenarios is in the
mechanism of the trapping.
In our model we considered multi-field trapping\footnote{We use the word
``multi-field'' in the sense that both inflaton $\phi_1$ and light scalar
$\phi_2$ are trapped at ESP}  
due to the confining force induced by the preheat field, instead of
      considering the thermal effective potential.
\end{itemize}

\end{document}